\documentclass[twocolumn,twoside]{article}

\usepackage{dsfont,amstext,times,multicol}

\usepackage{QC,ISTET99}

\def\bm#1{\mathchoice{\mbox{\boldmath{$\displaystyle #1$}}}%
{\mbox{\boldmath{$\textstyle #1$}}}%
{\mbox{\boldmath{$\scriptstyle #1$}}}%
{\mbox{\boldmath{$\scriptscriptstyle #1$}}}}

\def\C{\mathds{C}}
\def\F{\mathds{F}}

\def\openone{\mathds{1}}

\def\bra#1{\left<#1\right|}
\def\ket#1{\left|#1\right>}

\def\entspricht{\mathrel{\hat{=}}}

\def\sx{{\sigma_x}}
\def\sy{{\sigma_y}}
\def\sz{{\sigma_z}}

\def\wgt{\mathop{\rm wgt}\nolimits}
\def\tr{\mathop{\rm Tr}\nolimits}

\def\w{\omega}
\def\ww{\overline{\omega}}

\newtheorem{theorem}{Theorem}[section]

\newtheorem{coro}[theorem]{Corollary}
\newtheorem{const}[theorem]{Construction}
\newtheorem{defin}[theorem]{Definition}

\begin{document}
\newcounter{originalpage}
\setcounter{originalpage}{207}
\def\thepage{{\setcounter{originalpage}{\value{page}}\addtocounter{originalpage}{206}\sf\arabic{originalpage}}}

\title{Quantum BCH Codes}

\author{Markus Grassl and Thomas Beth}

\address{Institut f\"ur Algorithmen und Kognitive Systeme (IAKS)\\
Universit\"at Karlsruhe, Am Fasanengarten 5, 76\,128 Karlsruhe,
Germany\\
Email: {grassl@ira.uka.de}, {EISS\_Office@ira.uka.de}
}
\maketitle

\begin{abstract}
After a brief introduction to both quantum computation and quantum
error correction, we show how to construct quantum error-correcting
codes based on classical BCH codes. With these codes, decoding can
exploit additional information about the position of errors. This
error model---the quantum erasure channel---is discussed. Finally,
parameters of quantum BCH codes are provided.
\end{abstract}

\section{Introduction}
Motivated by the statement

``{\em BCH codes are among the best (classical) codes we know}''

(cited from Ch.~9, $\S$1, p.~258 of \cite{MS77}), we present the
translation of classical Bose-Chaudhuri-Hocquenghem (BCH) codes into
quantum quantum error-correcting codes. Without error correction, the
promising new field of quantum computing (see, e.\,g.,
\cite{Sho97,Gro96_search}) would be mainly of theoretical nature. A
main ingredient of quantum computation is constructive and destructive
interference of different computation paths which is only possible
when using quantum states. But on the other hand, any possible
computing device exploiting quantum mechanics has to cope with
uncontrollable interactions with the environment, e.\,g., single
photons. Quantum error-correcting codes help to actively reduce the
decoherence due to coupling to the environment.

\section{Background}\label{background}
\subsection{Quantum Registers}
Classically, information is often represented by bits. A single bit
takes either the value 0 or 1. In physical systems, 0 and 1 are
represented by two different states of the system. These could be two
different voltages, signals with two different frequencies, but also
states on the quantum mechanical level, e.\,g., ground state and
excited state of an electron of an atom or ion, the spin of a nucleus,
or the polarization of photons. In Dirac notation \cite{Dir58}, the
two states are written as
$$
\mbox{``0''}\entspricht
\ket{0}=\left(\begin{array}{c}1\\0\end{array}\right)\in\C^2 
$$
and
$$\mbox{``1''}\entspricht
\ket{1}=\left(\begin{array}{c}0\\1\end{array}\right)\in\C^2. 
$$

In quantum mechanics, the principle of superposition allows a system
to be simultaneously in different states. Mathematically, the state of
the basic unit of quantum information, a {\em quantum bit} (or short
{\em qubit}), is represented by the normalized linear combination 
$$
\ket{q}=\alpha\ket{0}+\beta\ket{1}\quad\mbox{where $\alpha,\beta\in\C$,
$|\alpha|^2+|\beta|^2=1$.}
$$
The normalization condition stems from the fact that when extracting
classical information from the quantum system by a measurement, the
values $0$ and $1$ occur with probability $|\alpha|^2$ and
$|\beta|^2$, resp.

Similar to classical registers, a quantum register is built by
combining several qubits. Mathematically, this corresponds to the
tensor product of two-dimensional vector spaces\footnote{In quantum
mechanics, the underlying structures are {\em Hilbert spaces}. We do
not stress this fact since here all vector spaces are finite
dimensional and thus complete w.\,r.\,t. the standard Hermitian inner
product.}. Hence the state of a quantum register of length $n$ can be
any normalized complex linear combination of the $2^n$ mutually
orthogonal basis states
$$
\ket{b_1}\otimes\ldots\otimes\ket{b_n}=:\ket{b_1\ldots
b_n}=\ket{\bm{b}}\mbox{ where $b_i\in\{0,1\}$.}
$$

\subsection{Quantum Gates}
The laws of quantum mechanics say that any transformation on quantum
systems is linear. Furthermore, in order to preserve the
normalization any operation has to be unitary. Let us first consider
operations involving only one qubit, i.\,e., one subsystem. Similar to
the classical $NOT$ gate, there is a quantum operation exchanging the
states $\ket{0}$ and $\ket{1}$ given by the matrix
$$
NOT:=\left(\begin{array}{cc}
0 & 1\\
1 & 0
\end{array}\right).
$$
But on a single qubit, there is not only this ``classical''
operation. Examples for non-classical operations on single qubits are
given by
\begin{equation}\label{hadamard}
H:=\frac{1}{\sqrt{2}}
\left(\begin{array}{rr}
1 &  1\\
1 & -1
\end{array}\right)
\quad\mbox{and}\quad
\sigma_z:=\left(
\begin{array}{rr}
1 & 0\\
0 & -1
\end{array}
\right).
\end{equation}
Besides single qubit operations, the so-called controlled $NOT$ gate
($CNOT$) plays an important r{\^o}le since any unitary operation on a
$2^n$-dimensional space can be implemented using only single qubit
operations and $CNOT$ gates (see \cite{BBC95}). As a classical gate,
the $CNOT$ gate corresponds to a gate with two inputs and two
outputs. One of the inputs is copied to the first output, the second
output is the $XOR$ of the inputs. The transformation matrix of the
$CNOT$ gate is given by:
$$
CNOT:=\left(\begin{array}{cccc}
1 & 0 & 0 & 0\\
0 & 1 & 0 & 0\\
0 & 0 & 0 & 1\\
0 & 0 & 1 & 0
\end{array}\right)
\qquad
\begin{picture}(80,30)(-10,5)
\put(-3,0){\makebox(0,0)[r]{$\ket{b}$}}
\put(-3,20){\makebox(0,0)[r]{$\ket{a}$}}
\put(0,0){\line(1,0){40}}
\put(0,20){\line(1,0){40}}
\put(20,0){\circle{10}}
\put(20,20){\line(0,-1){25}}
\put(20,20){\makebox(0,0){$\bullet$}}
\put(43,0){\makebox(0,0)[l]{$\ket{a\oplus b}$}}
\put(43,20){\makebox(0,0)[l]{$\ket{a}$}}
\end{picture}
$$
On the right hand side, the notation for the $CNOT$ gate as a quantum
circuit is given. Each of the horizontal lines ({\em wires})
corresponds to a qubit of the whole quantum register. The dot on the
upper wire indicates that the transformation on the lower qubit (the
target)---a $NOT$ gate---is only applied when the state of the upper
qubit (the control) is $\ket{1}$. More examples for quantum circuits
can be found in \cite{RoBe99}.

\section{Quantum Error Correction}
Classically, a major technique for protecting information against
channel errors is to add redundant information. The simplest example
is a repetition code where information is replicated by the sender. At
the receiver's end of the channel, the most likely information is
chosen based on comparing all received messages and taking a majority
vote.

This technique cannot be translated directly to quantum systems since
it is not possible to copy unknown quantum information ({\em
no-cloning theorem} \cite{WoZu82}), and comparison of quantum states
is only possible statistically. Nevertheless, quantum states can be
protected against errors. The main idea is to embed quantum
information represented by $k$ qubits into a larger Hilbert space of
$n$ qubits where $n>k$. 

For the construction of quantum error-correcting codes, we have to
model which types of errors occur during the transmission over a
quantum channel. This topic will be addressed next.

\subsection{Error Models}
\subsubsection{Open Quantum Systems}
We assume that our quantum system interacts with an environment which
is not or only partially accessible. Nevertheless, we can {\em model}
the interaction by a unitary transformation
$U_{\text{interaction}}=U_{\text{int}}$ on the Hilbert space formed by the
system and its environment. Assuming that there is no prior
entanglement of the system with the environment, the interaction
operator reads as
$$
\ket{\psi}_{\text{sys}}\otimes\ket{\Psi}_{\text{env}}
\longmapsto
U_{\text{interaction}}
\left(\ket{\psi}_{\text{sys}}\otimes\ket{\Psi}_{\text{env}}\right).
$$
After this interaction, the state need no longer be a tensor
product. Since we cannot control the environment, we have to discard
any information about the environment. This is mathematically
reflected by {\em tracing out the environment}:
\begin{eqnarray}
\rho_{\text{sys}}&=&\tr_{\text{env}}
\left(U_{\text{int}}\left(
\ket{\psi}\bra{\psi}_{\text{sys}}\otimes
\ket{\Psi}\bra{\Psi}_{\text{env}}\right)
U_{\text{int}}^\dagger\right)\nonumber\\
&=&\sum_j A_j
\left(\ket{\psi}\bra{\psi}_{\text{sys}}\right)
 A_j^\dagger.\label{KrausOp}
\end{eqnarray}
The state of our quantum system is now, in general, a mixed state
given by the density operator $\rho_{\text{sys}}$. One interpretation
of a mixed quantum state is that we have an ensemble of pure quantum
states chosen according to a probability distribution. In our case,
one can think of a measurement performed on the environment. Due to
entanglement with the system, this may lead to different states of the
system depending on the measurement outcome---but we do not know which
one since the result of the measurement is discarded.

In order to model a quantum channel, we make use of equation
(\ref{KrausOp}). The disturbed quantum state $\rho_{\text{sys}}$ can
be expressed only in terms of the initial state
$\ket{\psi}\bra{\psi}_{\text{sys}}$ of the system and
some interaction operators $A_j$ which completely specify the channel.

For a single qubit, i.\,e., a two-dimensional quantum system, the
operators $A_j$ can be chosen to be proportional to the identity
operator and the Pauli matrices
{\small\arraycolsep0.6\arraycolsep
$$
\sx:=\left(\begin{array}{rr}
0 & 1\\
1 & 0
\end{array}\right),\quad
\sy:=\left(\begin{array}{rr}
0 & -i\\
i &  0
\end{array}\right),\quad
\sz:=\left(\begin{array}{rr}
1 &  0\\
0 & -1
\end{array}\right)
$$}%
where ($i^2=-1$). Surprisingly, in order to correct an arbitrary error
it is sufficient to be able to correct any of these four errors.

For more than one qubit, an {\em error basis} can be formed by tensor
products of the one qubit interaction operators. A common assumption
is that the errors act independently on each qubit. Furthermore,
errors are assumed to be small, i.\,e., {\em near} identity (with
respect to a suitable operator norm). Then errors with a small number
of tensor factors different from identity are {\em more likely} than
those errors with a large number of tensor factors different from
identity.

\subsubsection{Depolarizing and Erasure Channel}
To illustrate the preceding, we consider two important quantum
channels. Over a depolarizing channel \cite{BDSW96}, quantum information is
transmitted undisturbed with probability $1-\varepsilon$, and it is
replaced by a completely randomized quantum state with probability
$\varepsilon$. In this case, equation (\ref{KrausOp}) reads
\begin{eqnarray*}
\rho_{\text{sys}}
&=&(1-\varepsilon)\cdot\ket{\psi}\bra{\psi}_{\text{sys}}
+\epsilon\cdot\openone\\
&=&\;(1-3/4\cdot\varepsilon)\;
    id \left(\ket{\psi}\bra{\psi}_{\text{sys}}\right) id\\
&&
+\varepsilon/4\sum_{j=x,y,z}\;\sigma_j
\left(\ket{\psi}\bra{\psi}_{\text{sys}}\right)
\sigma_j.
\end{eqnarray*}
A related quantum channel is the quantum erasure channel
\cite{GBP97}. Again, the quantum state is transmitted undisturbed with
probability $1-\varepsilon$. In case of an error, the quantum state is
replaced by a quantum state $\ket{e}$ that is orthogonal to all other
quantum states. Equation (\ref{KrausOp}) now reads
$$
\rho_{\text{sys}}
=(1-\varepsilon)\cdot \ket{\psi}\bra{\psi}_{\text{sys}}
+\epsilon\cdot\ket{e}\bra{e}.
$$
Similar to classical erasures, the state $\ket{e}$ indicates that an
error occurred, i.\,e., side-information about positions of errors is
available for the decoding process. Note that we have increased the
dimension of the Hilbert space of the system by one adding the state
$\ket{e}$. Alternatively, we may use any state of the original space
instead of $\ket{e}$ and describe the positions of errors by other
means.

\subsection{Code Constructions}
In this section, we will briefly describe several constructions of
quantum error-correcting codes based on classical linear
error-correcting codes. As discussed above, for qubit systems it is
sufficient to be able to correct any error that is a tensor product of
identity and Pauli matrices. The {\em weight} of such an error (or the
{\em number of errors}) is defined as the number of tensor factors
different from identity. Moreover, as $\sy=i\sx\sz$ we can restrict
ourselves to no-error, $\sx$-errors, $\sz$-errors, and combinations of
them. The operator $\sx$ interchanges the states $\ket{0}$ and
$\ket{1}$. Hence, it corresponds to a classical bit-flip error. The
operator $\sz$ changes the relative phase of $\ket{0}$ and $\ket{1}$
and has no classical counterpart. But the operator $\sz$ interchanges
the orthogonal states $(\ket{0}+\ket{1})/\sqrt{2}$ and
$(\ket{0}-\ket{1})/\sqrt{2}$, i.\,e., it acts as a bit-flip with
respect to this basis. Hence, the corresponding change of basis---the
Hadamard transform $H$ (see equation (\ref{hadamard}))---interchanges
bit-flip and phase-flip errors:
$$
H\sx H =\sz \quad\mbox{and}\quad H\sz H =\sx.
$$
In summary, this enables us to use certain classical linear binary
codes for the construction of quantum codes.

The following construction is due to \cite{Ste96:error,Ste96:multiple}
and \cite{CaSh96}. More details (and proofs) can also be found in
\cite{BeGr98,GrBe99:DPG}.
\begin{const}[Binary Codes]\label{constBinCode}\ \\
Let $C=[n,k,d]$ be a weakly self-dual linear binary code, i.\,e., $C$ is
contained in its dual $C^\bot=[n,n-k,d^\bot]$. Furthermore, let $\{\bm{w}_j:
0\le j\le 2^{n-2k}\}$ be a system of coset representatives of
$C^\bot/C$.

Then the $2^{n-2k}$ mutually orthogonal states
\begin{equation}\label{binQCodeState}
\ket{\psi_j}
  =\frac{1}{\sqrt{|C|}}\sum_{\bm{c}\in C}\ket{\bm{c}+\bm{w}_j}
\end{equation}
span a quantum error-correcting code ${\cal C}=[[n,n-2k]]$ of length $n$
and dimension $2^{n-2k}$ (The notation is similar to that for
classical linear block codes.) Based on classical decoding algorithms
for the code $C^\bot$, up to $(d^\bot-1)/2$ errors can be
corrected. Moreover, the code can correct errors up to weight
$(d'-1)/2$ where
\begin{equation}\label{truemindist2}
 d'=\min \{\wgt \bm{c}: \bm{c}\in C^\bot\setminus C\}\ge d^\bot.
\end{equation}
\end{const}
The outline of the decoding process is as follows: Any superposition
of code states $\ket{\psi_j}$ is a superposition of quantum states
corresponding to codewords of the dual code $C^\bot$. A (correctable)
bit-flip error takes the superposition of codewords into a
superposition of the corresponding coset. Similar to classical
decoding algorithms, this coset can be identified by computing an
error syndrome using auxiliary qubits. Measuring this syndrome reveals
information about the error, but not about the original
superposition. After correction of the bit-flip errors, a Hadamard
transform turns the remaining phase-flip errors into sign-flip
errors. The Hadamard transform changes the code state
(\ref{binQCodeState}) into
\begin{equation}\label{binQCodeStateHad}
H^{\otimes n}\ket{\psi_j}
  =\frac{1}{\sqrt{|C^\bot|}}\sum_{\bm{c}\in C^\bot}
(-1)^{\bm{c}\cdot\bm{w}_j}\ket{\bm{c}}.
\end{equation}
Here $\bm{c}\cdot\bm{w}_j$ is the standard inner product $\bm{x}\cdot
\bm{y}=\sum_i x_i y_i$. Again, any superposition of states
(\ref{binQCodeStateHad}) is a superposition of quantum states
corresponding to codewords of the dual code $C^\bot$. Hence the errors
can be corrected in the same manner. The last step is another Hadamard
transform returning to the original basis.

A generalization of this construction was given in
\cite{Got96_hamming} and \cite{CRSS98}. It is based on the algebraic
properties of the group generated by tensor products of Pauli matrices
(see also \cite{BeGr98}). Here we will only present the prerequisites
and the parameters of the resulting quantum codes. Furthermore, we
restrict ourselves to linear codes (in contrast to {\em additive}
codes).
\begin{const}[Quaternary Codes]\label{constGF4Code}\ \\
By $\overline{x}$ we denote the conjugation $x\mapsto
x^2=:\overline{x}$ in the field
$\F_4=GF(4)=\{0,1,\w,\ww=\w^2=\w+1\}$. Furthermore, for a linear space
$C\le \F_4^n$, by $C^*$ we denote the linear space that is orthogonal
with respect to the inner product
$\bm{x}\cdot\bm{y}:=\sum_{j}\overline{x_j}y_j$.

Let $C=[n,k,d]$ be a self-orthogonal linear quaternary code, i.\,e., $C$
is contained in $C^*=[n,n-k,d^*]$.

Then a quantum error-correcting code ${\cal C}=[[n,n-2k]]$ of length
$n$ and dimension $2^{n-2k}$ exists. Based on classical decoding
algorithms for the code $C^*$, up to $(d^*-1)/2$ errors can be
corrected. Moreover, the code can correct errors up to weight
$(d'-1)/2$ where
\begin{equation}\label{truemindist4}
 d'=\min \{\wgt \bm{c}: \bm{c}\in C^*\setminus C\}\ge d^*.
\end{equation}
\end{const}
Note that $C^\bot$ and $C^*$ are related by conjugation and thus
$d^*=d^\bot$.

Recently, it has been shown how to use linear codes over any finite
field of characteristic two, i.\,e., fields $\F_{2^\ell}$ with $2^\ell$
elements for the construction of quantum error-correcting codes
\cite{GGB99}. Again, we only present the main parameters of the
construction.
\begin{const}[Codes from Extension Fields]\label{constExtCode}\ \\
Let $C=[n,k,d]$ be a weakly self-dual code over $\F_{2^\ell}$, i.\,e.,
$C$ is contained in its dual $C^\bot=[n,n-k,d^\bot]$ (with respect to
the standard inner product). Furthermore, let $B$ be a self-dual basis
of $\F_{2^\ell}$ over $\F_2$.

Expanding each element of $\F_{2^\ell}$ with respect to the basis $B$
yields a weakly self-dual linear binary code $C_2=[\ell n,\ell
k,d_2\ge d]$. Its dual $C_2^\bot=[\ell n,\ell(n-k),d_2^\bot\ge
d^\bot]$ is obtained in the same manner.

Based on the classical codes $C_2$ and $C_2^\bot$, a quantum
error-correcting code can be obtained using Construction
\ref{constBinCode}. The resulting quantum code can be decoded as a
binary code or as a code over the field $\F_{2^\ell}$. In the latter
case, $\ell$ qubits are grouped into one block, and errors can be
corrected if they are restricted to up to $(d^\bot-1)/2$ blocks.
\end{const}\vspace*{-0.95mm}

\section{Quantum BCH Codes}
The quantum version of binary BCH codes was introduced in
\cite{GBP97}. In \cite{CRSS98}, the term quantum BCH code was used for
quaternary quantum BCH codes (see Construction \ref{constGF4Code}). In
the context of \cite{GBP97}, for the quantum erasure channel, it is
important to use codes that allow the use of the side-information on
the positions of the errors provided by the channel. For BCH codes, a
variety of such decoding algorithms exists. Being cyclic codes, BCH
codes allow also decoding based on spectral techniques. This is in
particular true for Reed-Solomon (RS) codes where no field extension
is needed to implement the Fourier transform. The quantum version of
RS codes and their spectral decoding is discussed in
\cite{GGB99}. Another technique for encoding and decoding cyclic codes
is based on linear shift registers (see \cite{GrBe98}).

In the sequel, we focus on the definition and the computation of the
parameters of quantum BCH codes, supplemented by examples in
Section~\ref{Examples}. A good reference for the theory of classical
error-correcting codes is \cite{MS77}. All theorems below can be found
in a similar version in \cite{GBP97} and \cite{CRSS98}, we will omit
the proofs.

\begin{defin}[${\cal Q}$BCH Codes]\ \\
A quantum BCH code (${\cal Q}$BCH code) is a quantum error-correcting
code that is derived from a classical, weakly self-dual (respectively
self-orthogonal) BCH code using Construction \ref{constBinCode},
\ref{constGF4Code}, or \ref{constExtCode}.
\end{defin}

Usually, BCH codes are specified by the {\em zero sets}, i.\,e., the
exponents of the roots $\alpha^z$ of their generator polynomial
$g(X)|X^n-1$ where $\alpha$ is a primitive $n$-th root of unity. For
a BCH code over the field $\F_q$, the zero set is a union of
cyclotomic cosets modulo $n$ closed under
multiplication by $q$, i.\,e.,
$$
{\cal Z}_C=\bigcup_z C_z\quad\mbox{where $C_z=\{q^i z\bmod n: i\ge 0\}$.}
$$

The zero sets of a code and its dual are related as follows.
\begin{theorem}
Let ${\cal Z}_C$ denote the zero set of a BCH code $C$ over the field
$\F_q$, i.\,e., the generator polynomial of $C$ is given by
$$
g(X)=\prod_{z\in{\cal Z}_C}(X-\alpha^z).
$$
Then the generator polynomial of the dual code $C^\bot$ is given by
$$
h(X)=\prod_{z\in\{0,\ldots,n-1\}\setminus{\cal Z}_C}(X-\alpha^{-z}),
$$
i.\,e., the zero set of the dual code is given by 
$$
{\cal Z}_{C^\bot}=\left\{-z\bmod n: z\in \{0,\ldots,n-1\}\setminus{\cal
Z}_C\right\}.
$$
For codes over $\F_4$, the generator polynomial of the orthogonal code
$C^*$ is given by 
$$ 
h(X)=\prod_{z\in\{0,\ldots,n-1\}\setminus{\cal Z}_C}(X-\alpha^{-2z}),
$$
i.\,e., the zero set of the orthogonal code is given by 
$$
{\cal Z}_{C^*}=\left\{-2z \bmod n: z\in \{0,\ldots,n-1\}\setminus{\cal
Z}_C\right\}.
$$
\end{theorem}

\begin{coro}
A BCH code is weakly self-dual if and only if 
${\cal Z}_{C^\bot}\subseteq {\cal Z}_C$ or, equivalently, 
$$
\forall z: \left(z\in {\cal Z}_{C^\bot}\Rightarrow (-z \bmod n)
  \notin {\cal Z}_C\right).
$$
A BCH code over $\F_4$ is self-orthogonal if and only if
${\cal Z}_{C^*}\subseteq {\cal Z}_C$ or, equivalently, 
$$
\forall z: \left(z\in {\cal Z}_{C^*}\Rightarrow (-2^{-1}z \bmod n)
  \notin {\cal Z}_C\right).
$$
\end{coro}

A lower bound for the minimum distance of a BCH code---and in turn for
the corresponding ${\cal Q}BCH$ code---can be derived from its zero
set.
\begin{theorem}[BCH bound]
If the zero set ${\cal Z}_{C^\bot}$ of the dual of a weakly self-dual
BCH code $C$ contains $d_{\text{BCH}}-1$ consecutive numbers, i.\,e.,
\begin{equation}\label{designed_distance}
\bigcup_{z=z_0}^{z_0+d_{\text{BCH}}-2} C_z\subseteq {\cal Z}_{C^\bot},
\end{equation}
then the minimum distance $d^\bot$ of $C^\bot$ is at least
$d_{\text{BCH}}$.
\end{theorem}
On the other hand, if a BCH code is specified by the left hand side of
equation (\ref{designed_distance}), $d_{\text{BCH}}$ is called the {\em
designed distance}. 

The actual minimum distance of a BCH code may be larger than
$d_{\text{BCH}}$. This yields another lower bound for the error
correcting capability of the ${\cal Q}BCH$ code.
\begin{theorem}[Code bound]
The minimum distance of a ${\cal Q}BCH$ code is at least the minimum
distance $d^\bot$ of the dual $C^\bot=[n,n-k,d^\bot]$ of the
underlying BCH code.
\end{theorem}

According to equations (\ref{truemindist2}) and (\ref{truemindist4}),
the true minimum distance of a ${\cal Q}BCH$ code may be even larger,
see the examples in the next section.

\section{Examples}\label{Examples}
Finally, we present the main results of this paper. Using the computer
algebra system {\sf MAGMA} \cite{Magma}, we have computed the
parameters for ${\cal Q}$BCH codes derived from classical BCH codes
over various fields (see Tables~\ref{tableGF2}--\ref{tableGF64}).

In Table~\ref{tableGF2} parameters of binary ${\cal Q}$BCH codes are
given. A noteable code is the one with parameters ${\cal
C}=[[49,1,9]]$. The corresponding BCH code is $C^\bot=[49, 25, 4]$ and
$C=[49,24,4]$ is the even weight subcode of $C^\bot$. Therefore,
$d'=\min \{\wgt \bm{c}: \bm{c}\in C^\bot\setminus C\}$ must be
odd. Computing the weight distribution of $C^\bot$, we obtain
$d'=9$. 

Similarly, for the code ${\cal C}=[[89,1,17]]$ the BCH bound yields
$d'\ge 7$, whereas the actual minimum distance of the BCH code
$C^\bot$ is $d^\bot=12$. Again, $C=[89,44,12]$ is the even weight
subcode of $C^\bot$, hence $d'\ge 13$ and $d'$ is odd. Sampling
codewords at random, we find $d'\le 17$. Moreover, using {\sf MAGMA}
we were able to show that indeed $d'=17$.

Quaternary ${\cal Q}$BCH codes are listed in
Table~\ref{tableGF4}. Here are the codes ${\cal C}=[[25, 1,9]]$ and
${\cal C}=[[35, 1,9]]$ of special interest. For the first code, the
BCH bound yields $d'\ge 4$, the minimum distance of both $C$ and $C^*$
is $d=8$, but the minimum distance of the quantum code is $d'=9$. For 
${\cal C}=[[35, 1,9]]$, we obtain $d_{\text{BCH}}=5$, $d=8$, and
$d'=9$.

Finally, in Tables~\ref{tableGF8}--\ref{tableGF64} we present ${\cal
Q}$BCH codes constructed from BCH codes over fields of size $8$, $16$,
$32$, and $64$. The corresponding binary codes are obtained by
expanding each element of the extension field with respect to a fixed
self-dual basis. For these codes, we have listed both the minimum
distance $d_2$ as binary code and the minimum distance $d_q$ as code
over the field $\F_q$ which is relevant for blockwise decoding.

\begin{table}[hbt]
\begin{center}
\begin{minipage}{0.9\hsize}\small
\parskip5pt plus 3pt
\begin{multicols}{3}
$[[ 7, 1, 3]]$

$[[15, 7, 3]]$

$[[21, 3, 5]]$\\
$[[21, 9, 3]]$\\
$[[21,15, 2]]$

$[[23, 1, 7]]$

$[[31, 1, 7]]$\\
$[[31,11, 5]]$\\
$[[31,21, 3]]$

$[[35, 5, 6]]$\\
$[[35,11, 3]]$\\
$[[35,29, 2]]$

$[[39,15, 3]]$

$[[45,13, 5]]$\\
$[[45,21, 3]]$\\
$[[45,37, 2]]$

$[[47, 1,11]]$

$[[49, 1, 9]]$\footnote{The code bound yields only $d\ge 4$.}\\ 
$[[49, 7, 3]]$\\
$[[49,43, 2]]$

$[[51,35, 3]]$

$[[55,15, 5]]$

$[[63,27, 7]]$\\
$[[63,39, 5]]$\\
$[[63,45, 4]]$\\
$[[63,51, 3]]$\\
$[[63,57, 2]]$

$[[69, 3,11]]$\\
$[[69,25, 3]]$\\
$[[69,47, 2]]$

$[[71, 1,11]]$

$[[73,19, 9]]$\\
$[[73,37, 6]]$\\
$[[73,55, 3]]$

$[[75,35, 3]]$\\
$[[75,67, 2]]$

$[[77,11, 6]]$\\
$[[77,17, 3]]$\\
$[[77,71, 2]]$

$[[79, 1,15]]$

$[[85,53, 5]]$\\
$[[85,69, 3]]$

$[[87,31, 3]]$

$[[89, 1,17]]$\footnote{The code bound yields only $d\ge 12$.}\\ 
$[[89,23,11]]$\\
$[[89,45, 7]]$\\
$[[89,67, 4]]$

$[[91,43, 7]]$\\
$[[91,67, 3]]$\\
$[[91,85, 2]]$

$[[93,13,12]]$\\
$[[93,23, 9]]$\\
$[[93,33, 8]]$\\
$[[93,43, 7]]$\\
$[[93,63, 5]]$\\
$[[93,73, 3]]$\\
$[[93,83, 2]]$

$[[95,23, 5]]$

$[[103, 1,19]]$

$[[105,37,9]]$\\
$[[105,45,7]]$\\
$[[105,61,5]]$\\
$[[105,75,4]]$\\
$[[105,91,3]]$\\
$[[105,99,2]]$

$[[111,39,3]]$

$[[115, 5,14]]$\\
$[[115,27, 5]]$\\
$[[115,93, 2]]]$  
	    
$[[117,45, 9]]$\\
$[[117,69, 7]]$\\
$[[117,93, 3]]$

$[[119, 23, 7]]$\\
$[[119, 65, 6]]$\\
$[[119, 71, 3]]$\\
$[[119,113, 2]]$

$[[123,83, 3]]$

$[[127,1, 19]]$\\
$[[127,15,16]]$\\
$[[127,29,15]]$\\
$[[127,43,13]]$\\
$[[127,57,11]]$\\
$[[127,71,9]]$\\
$[[127,85,7]]$\\
$[[127,99,5]]$\\
$[[127,113,3]]$
\end{multicols}
\end{minipage}
\end{center}

\caption{Parameters of some binary ${\cal Q}$BCH codes
given in the form $[[n,k,d]]$.
\label{tableGF2}}
\end{table}

\vspace*{-5mm}
\section{Acknowledgments}
The authors would like to thank Rainer Steinwandt
for his comments. \hskip0pt plus2pt 
This work was supported by {\em Deutsche Forschungsgemeinschaft (DFG),
\hskip0pt plus2pt Schwerpunktprogramm Quanten-Informationsverarbeitung
(SPP 1078), \hskip0pt plus2pt Projekt AQUA (Be~887/13-1)}.

\begin{table}[hbt]
\begin{center}
\begin{minipage}{0.9\hsize}\small
\parskip5pt plus 3pt
\begin{multicols}{3}
$[[5, 1,3]]$

$[[7, 1,3]]$

$[[13, 1,5]]$

$[[15, 3,5]]$\\
$[[15, 7,3]]$\\
$[[15,11,2]]$

$[[17, 1,7]]$\\
$[[17, 9,4]]$

$[[21, 3,5]]$\\
$[[21, 9,3]]$\\
$[[21,15,2]]$

$[[23, 1,7]]$

$[[25, 1,9]]$\footnote{The code bound yields only $d\ge 4$.}\\
$[[25, 5,3]]$\\
$[[25,21,2]]$

$[[29, 1,11]]$

$[[31, 1,7]]$\\
$[[31,11,5]]$\\
$[[31,21,3]]$

$[[35, 1,9]]$\footnote{The code bound yields only $d\ge 8$.}\\
$[[35,13,7]]$\\
$[[35,25,4]]$\\
$[[35,31,2]]$

$[[37, 1,11]]$

$[[39,3,9]]$\\
$[[39,15,3]]$\\
$[[39,27,2]]$

$[[41,1,11]]$\\
$[[41,21,6]]$

$[[45,17,5]]$\\
$[[45,29,3]]$\\
$[[45,41,2]]$

$[[47,1,11]]$

$[[49, 1,9]]$\footnote{The code bound yields only $d\ge 4$.}\\
$[[49, 7,3]]$\\
$[[49,43,2]]$

$[[51, 3,11]]$\\
$[[51,19,9]]$\\
$[[51,27,6]]$\\
$[[51,35,3]]$\\
$[[51,43,2]]$

$[[53, 1,15]]$

$[[55,31, 5]]$\\ 
$[[55,35, 3]]$\\
$[[55,51, 2]]$

$[[61,1,17]]$
\end{multicols}
\end{minipage}
\end{center}

\caption{Parameters of some quaternary ${\cal Q}$BCH codes given in the form
$[[n,k,d]]$.\label{tableGF4}}
\end{table}
\begin{table}[hbt]
\rule{0pt}{0pt}\vskip0mm minus 5mm

\begin{center}
\begin{minipage}{0.9\hsize}
\begin{multicols}{3}\parskip5pt plus 3pt
\small
$[[ 21,15,2|2]]$\\
$[[ 21, 9,3|3]]$

$[[ 45,21,3|3]]$

$[[ 63,27,6|5]]$\\
$[[ 63,21,5|5]]$\\
$[[ 63,33,4|4]]$\\
$[[ 63,45,3|3]]$\\
$[[ 63,57,2|2]]$\newpage

$[[ 69, 3,7|7]]$

$[[ 93, 3,7|7]]$\\
$[[ 93,33,5|5]]$\\
$[[ 93,63,3|3]]$

$[[105,27,6|5]]$\\
$[[105,51,5|4]]$\\
$[[105,75,3|3]]$\\
$[[105,99,2|2]]$

$[[117,93,3|3]]$\\
$[[117,69,3|3]]$

$[[135, 87,5|5]]$\\
$[[135,111,3|3]]$
\end{multicols}
\end{minipage}
\end{center}

\caption{Parameters of some binary quantum codes derived from BCH
codes over $\F_8$ given in the form $[[n,k,d_2|d_8]]$. Binary expansion
with respect to the self-dual basis $B_8=(u^3,u^6,u^5)$
where $u^3=u+1$.\label{tableGF8}}
\end{table}
\rule{0pt}{0pt}\vskip-15mm plus 5mm

\begin{table}[hbt]
\begin{center}
\begin{minipage}{0.9\hsize}\small
\parskip5pt plus 3pt
\begin{multicols}{3}

$[[20,12,2|2]]$

$[[28,4,3|3]] $

$[[36, 4,4|3]]$\\
$[[36,28,2|2]]$

$[[44,4,6|5]] $

$[[52, 4,7|6]]$\\
$[[52,28,4|4]]$

$[[60,12,8|7]]$\\
$[[60,28,6|5]]$\\
$[[60,36,4|4]]$\\
$[[60,44,3|3]]$\\
$[[60,52,2|2]]$

$[[76,4,7|7]]$

$[[84, 4,6|6]]$\\
$[[84,28,5|5]]$\\
$[[84,36,4|3]]$\\
$[[84,52,3|3]]$\\
$[[84,76,2|2]]$

$[[92,4,7|7]]$

$[[100,12,9|6]]$\footnote{The code bound yields only $d_2\ge 8$
and $d_{16}\ge 4$.}\\
$[[100,52,4|3]]$\\
$[[100,92,2|2]]$

$[[108,28,4|4]]$\footnote{The code bound yields $d_2\ge 4$ and
$d_{16}\ge 3$. In contrast to similar cases, $C_2^\bot\setminus C_2$
contains words of minimum weight, hence $d_2=4$.}\\
$[[108,100,2|2]]$

$[[116,4,15|11]]$\footnote{The set $C^\bot\setminus C$ contains words of
minimum weight, hence the true minimum distance meets the code bound.}\\ 
$[[116,60,7|6]]$
\end{multicols}
\end{minipage}
\end{center}

\caption{Parameters of some binary quantum codes derived from BCH
codes over $\F_{16}$ given in the form $[[n,k,d_2|d_{16}]]$. Binary expansion
with respect to the self-dual basis $B_{16}=(v^3, v^7, v^{13}, v^{12})$
where $v^4=v+1$.\label{tableGF16}}
\end{table}

\begin{table}[ht]
\begin{center}
\begin{minipage}{0.9\hsize}
\begin{multicols}{3}\parskip5pt plus 3pt
\small
$[[35,5,3|3]]$

$[[75,35,3|3]]$

$[[105,15,5|5]]$\\
$[[105,45,3|3]]$\\
$[[105,75,2|2]]$

$[[115,5,7|7]]$

$[[175,25,6|6]]$\\
$[[175,55,3|3]]$\\
$[[175,145,2|2]]$
\end{multicols}
\end{minipage}
\end{center}

\caption{Parameters of some binary quantum codes derived from BCH
codes over $\F_{32}$ given in the form $[[n,k,d_2|d_{32}]]$. Binary expansion
with respect to the self-dual basis $B_{32}=(w^9, w^{18}, w^5, w^{10}, w^{20})$
where $w^5=w^2+1$.\label{tableGF32}}
\vskip10mm
\rlap{\hbox to \textwidth{\hrulefill}}
\end{table}

\newpage
\ifnum \value{page} > 5\else\ \newline\fi

\begin{table}[t]
\begin{center}
\begin{minipage}{0.9\hsize}
\begin{multicols}{3}\parskip5pt plus 3pt
\small
$[[42,18,3|3]]$\\
$[[42,30,2|2]]$

$[[54,18,6|4]]$\\
$[[54,30,4|3]]$\\
$[[54,42,2|2]]$

$[[66,6,6|5]]$

$[[90,54,3|3]]$\\
$[[90,78,2|2]]$

$[[114,42,8|6]]$\\
$[[114,78,6|4]]$

$[[126,54,8|7]]$\\
$[[126,78,6|5]]$\\
$[[126,90,4|4]]$\\
$[[126,102,3|3]]$\\
$[[126,114,2|2]]$
\end{multicols}
\end{minipage}
\end{center}

\caption{Parameters of some binary quantum codes derived from BCH
codes over $\F_{64}$ given in the form $[[n,k,d_2|d_{64}]]$. Binary expansion
with respect to the self-dual basis $B_{64}=(z^{12},z^{24},z^{48},z^{33},z^3,z^6)$
where $z^6=z^4+z^3+z+1$.\label{tableGF64}}
\vskip12mm
\end{table}

\end{document}